\documentclass[aps,prb,twocolumn,amsmath,amssymb,floatfix,superscriptaddress,longbibliography]{revtex4-2}
\usepackage{natbib}
\usepackage{graphicx,xcolor,multirow,bm,hyperref}
\usepackage{float}
\usepackage[ignoreunlbld,norefs,nocites]{refcheck}
\begin{document}

\title{Elastic properties of transition metal dichalcogenides} 
\author{S. Azadi}
\email{sam.azadi@manchester.ac.uk}
\affiliation{Department of Physics and Astronomy, University of Manchester, Oxford Road, Manchester M13 9PL, United Kingdom}
\author{A. Azhar}
\affiliation{Department of Physics and Astronomy, University of Manchester, Oxford Road, Manchester M13 9PL, United Kingdom}
\affiliation{Physics Study Program, Faculty of Science and Technology, Syarif Hidayatullah State Islamic University Jakarta, Tangerang Selatan 15412, Indonesia}
\author{R. V. Belosludov}
\affiliation{Institute for Materials Research, Tohoku University, Sendai 980-08577, Japan}
\author{T. D. K\"{u}hne}
\affiliation{Center for Advanced Systems Understanding, Untermarkt 20, D-02826 G\"orlitz, Germany}
\affiliation{Helmholtz Zentrum Dresden-Rossendorf, Bautzner Landstra{\ss}e 400, D-01328 Dresden, Germany}
\affiliation{TU Dresden, Institute of Artificial Intelligence, Chair of Computational System Sciences, N\"othnitzer Stra{\ss}e 46 D-01187 Dresden, Germany}
\author{M.\ S.\ Bahramy}
\affiliation{Department of Physics and Astronomy, University of Manchester, Oxford Road, Manchester M13 9PL, United Kingdom}

\begin{abstract}
We present a comprehensive first-principles study of the structural and elastic properties of 2H-MX$_2$ transition metal dichalcogenides (TMDs) (M = W, Mo, Ta, Nb; X = S, Se). Using density functional theory with various van der Waals exchange-correlation functionals, we systematically investigate the influence of nonlocal interactions on lattice parameters, elastic constants, and mechanical moduli. Our results reveal a fundamental distinction between semiconducting and metallic TMDs: metallic compounds exhibit larger in-plane lattice parameters and reduced interlayer spacing, consistent with their bonding characteristics. We find that metallic TMDs display significantly lower in-plane stiffness and shear modulus compared to their semiconducting counterparts. We discuss this behavior in the context of the observed charge density waves. In addition, we establish clear trends in the bulk, Young's, and shear moduli, demonstrating the role of atomic number and chemical composition in determining mechanical stability.
\end{abstract}

\maketitle

\section{Introduction}
 Layered transition metal dichalcogenides (TMDCs)\cite{Frindt,Joensen} are materials with weak van der Wall interactions (vdW) between planes and strong atomic bonding in the plane\cite{Mattheis,Wilson,Ayari,QWang}.  The electronic and optical properties of the TMDs can be tuned by applying strain\cite{WSYun,Johari,Rice,CLee, JLi}. As a prototype of TMDs,  MoS$_2$ has been widely studied due to its special mechanical, electronic, and optical properties. Raman and photoluminescence spectroscopy for MoS$_2$ indicate that the monolayer Raman mode and the nature of the band gap are affected by an applied strain\cite{CRZhu}. The elastic and mechanical properties of ultra thin freely suspended MoS$_2$ were studied and the Young's modulus of 0.35$\pm$0.02 (TPa) was determined \cite{Bertolazzi,RCCooper}.  A single layer of MoS$_2$ can be strained at least up to 10\%\cite{Bertolazzi} which is comparable to graphene. An applied strain reduces direct and indirect band gaps. The reduction in the band gap is approximately linear with strain \cite{KHe}. Because of a crossover from the direct to the indirect band gap that results from an applied strain, the emission efficiency of single-layers is expected to decrease for highly strained systems. This feature allows for mechanical tuning of the electronic properties and the possibility of fabrication of flexible electronics.

The structural, electronic, elastic, and dynamic properties of TMDs have been studied by applying density functional theory (DFT) and mainly using generalized gradient approximations \cite{JLi, KHe, Molina,Lorenz,YCai,WHuang,HPKomsa,Mirhosseini}. Non-local van der Walls (vdW) interactions play an important role in TMDs \cite{Peelaers,Bjrkman,Hummer}. Conventional DFT studies with local or semilocal XC functionals could fail in providing accurate results for TMDs. The reason for this failure is that the dominant part of the stabilization energy in TMDs comes from the dispersion energy, which is not properly taken into account in standard XC functionals.

\begin{figure}
\includegraphics[width=0.45\textwidth]{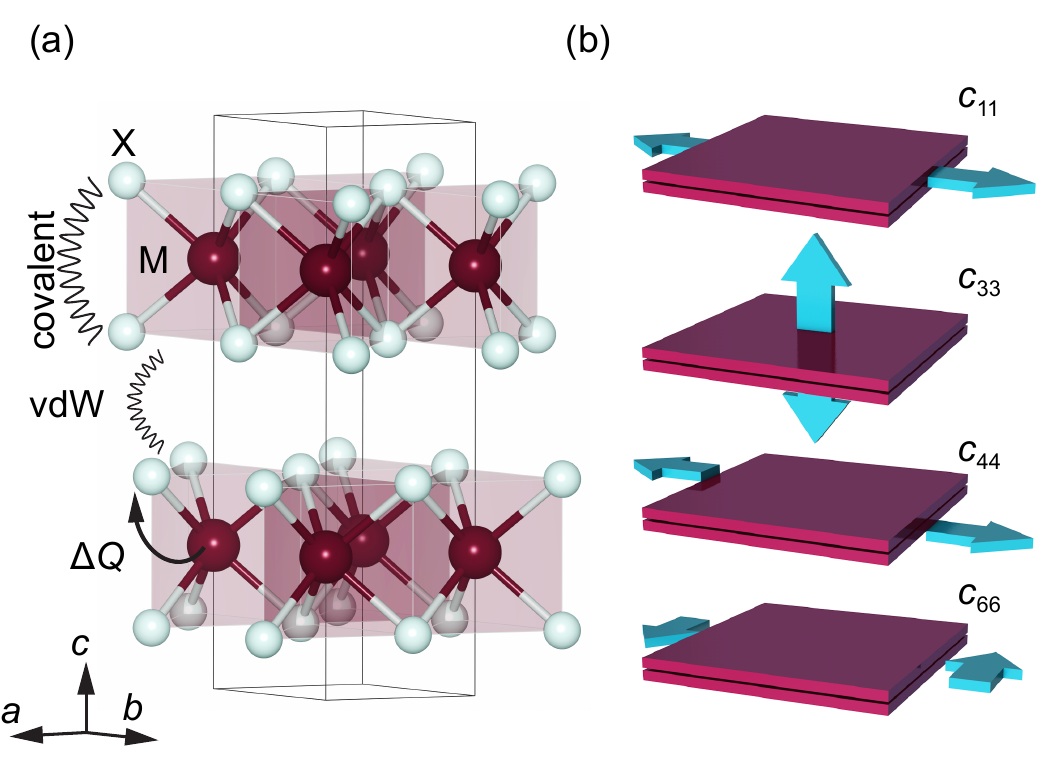}
\caption{\label{structure} (a) Lattice structure of MX$_2$, with M = W, Mo, Ta, and Nb, and X = S and Se. $\Delta Q$ denotes the charge transferred from M cation to X anion. The X ions can also form covalent bonding within each layer or interact via vdW forces between the adjacent layers. (b) Elastic constants representing the response of the MX$_2$ lattice to in-plane ($c_{11}$) and out-of-plane ($c_{33}$) stress as well as axial ($c_{44}$) and diagonal ($c_{66}$) shear deformations.} 
\end{figure}
In this work, we study the elastic properties of layered 2H-MX$_2$ with M = W, Mo, Ta and Nb and X = S and Se. We apply DFT method with well-tested vdW functionals as well as one semilocal exchange-correlation (XC) functional. It has been widely accepted that the XC approximation affects the final DFT results \cite{PRB13,JCP16,PCCP17}.  Hence, it is important to systematically study the influence of the XC approximation and compare it to the experiment. More importantly, to discover the potential applications of TMDs in flexible devices, an accurate characterization of their elastic properties is crucial. Using the DFT results and comparing the structural properties of metallic and semiconducting TMDs, we provide a general trend for TMDs that is independent of the XC approximation.

\section{Theoretical and computational details}\label{TCD}
 Our DFT calculations were performed within the pseudopotential and plane-wave approach implemented in the Quantum ESPRESSO suite of programs\cite{QE1,QE2}. We used six different vdW functionals\cite{vdwreview} of vdW-DF1\cite{vdw10,vdw11,vdw007},  vdW-DF2\cite{vdw2}, vdW-rVV10\cite{rvv1,rvv2,rvv3}, vdW-cx\cite{cx}, vdW-DF1-c09 and vdW-DF2-c09\cite{c09}. To compare with vdW results, the PBEsol\cite{PBEsol} XC functional is applied too. We used ultrasoft GBRV (US) PPs\cite{gbrv} and a basis set of plane waves with an energy cutoff of 80 Ry, while the charge density cutoff was set to 800 Ry. For metallic systems, the smearing method was applied using the Methfessel-Paxton\cite{MP} approach with a smearing parameter of 0.005 Ry. We considered $\bf k$-point mesh of $16\times16\times4$ for semiconductor systems and $24\times24\times8$ for metallic structures.

The elastic constants are simulated using a method based on the numerical differentiation of stress that has been applied to different crystals\cite{DalCorso}.  For TMDs with hexagonal symmetry, there are two independent lattice parameters of $a$ and $c$. To optimize the lattice parameters, $E(a, \frac{c}{a})$ was calculated on a uniform $8\times8$ grid of parameters in the space of $a$ and $\frac{c}{a}$. The step between the $a$ lattice constant was 0.05 a.u., whereas for $\frac{c}{a}$ it was 0.02 a.u. The total ground state energies were then fitted on a quartic polynomial. Before each counter calculation $E(a, \frac{c}{a})$, an initial primitive cell of the experimental structure was relaxed to zero pressure\cite{Sabatini}. The quasi-Newton algorithm was used for internal coordinates optimization, with convergence thresholds on the total energy and forces of 0.01 mRy and 0.1 mRy/Bohr, respectively. The final optimized structure was used to calculate the elastic properties.  We calculated the non-zero components of the stress matrix for a set of strains and obtained the elastic constants from the first numerical derivative of the stress with respect to strain. The number of strained configurations used to calculate each derivative was set to six and the strain value interval was selected as 0.005. 

To understand the difference between the results obtained by different vdW-XC functionals, we briefly describe the form of these functionals.  The vdW-XC functionals used in this work can be separated into local and non-local terms: $E_\text{XC}^\text{vdW} = E_{\text{XC},l}+E_{c,nl}$. The Slater exchange and Perdew-Wang (PW)\cite{pw} correlation functionals are used in vdW-DF1 and vdW-DF2, which means that the correlation energy is approximated by the local density approximation (LDA)\cite{LDA}. In vdW-DF1, the gradient correction on exchange energy uses the revised version of PBE\cite{PBE,rpb}, while vdW-DF2 uses an optimized version of PW86\cite{pw86}, which is named PW86R\cite{pw86r}. These functionals use different kernel for non-local energy term which accounts approximately for the non-local electron correlation effects. The nonlocal term is obtained using double space integration, which represents an improvement compared to local or semilocal functionals, especially in the case of layered structures\cite{Rydberg03}.  The vdW-rVV functional can be expressed as $E_\text{XC}^\text{vdW-rVV} = E_c^\text{PBE} + E_x^\text{PW86R} + E_{c,nl}^\text{rVV}$.

Phonon spectra were simulated using density functional perturbation theory (DFPT) \cite{Baroni}. We performed phonon band structure calculations along the $\Gamma$–$M$ direction using various XC functionals to evaluate their impact on CDW formation. The simulations employed a $24\times24\times8$ k-point grid and an $8\times8\times4$ q-point grid, with a plane-wave cutoff of 60 Ry. A small Methfessel–Paxton broadening of 0.005 Ry was applied to capture CDW formation and control electronic temperature which has a considerable impact on the dynamic of CDW and phonon spectra in metallic systems\cite{FePRB24,AuNat25,PRB25}.

\section{Results and discussion}\label{RD}
\subsection{Lattice parameters}
The atomic arrangement of the 2H-MX$_2$ compounds is illustrated in Figure~\ref{structure}-(a), highlighting key bonding characteristics and charge transfer effects. The unit cell consists of two MX$2$ layers stacked along the crystalline $c$-axis. Within each layer, a transition metal (M) atom is coordinated by six chalcogen (X) atoms in a trigonal prismatic configuration, exhibiting $D_{3h}$ symmetry. The adjacent layers are rotated 180$^{\circ}$ relative to each other. The intralayer M-X bonds are predominantly ionic, ensuring structural stability, whereas the interlayer interactions are governed by weak van der Waals forces. The degree of ionic charge transfer, $\Delta Q$, from the metal to the chalcogen plays a crucial role in determining the electronic and mechanical properties of the material. In semiconducting MX$_2$ compounds, M$^{4+}$ cations (e.g., Mo and W) fully transfer their charge to the X$^{2-}$ anions, resulting in a bulk semiconducting behavior. In contrast, in metallic MX$_2$, where M$^{3+}$ cations (e.g., Nb and Ta) exhibit incomplete charge transfer, additional covalent bonding emerges between X atoms, altering structural and electronic properties \cite{QWang}.

\begin{figure}
\includegraphics[width=0.40\textwidth]{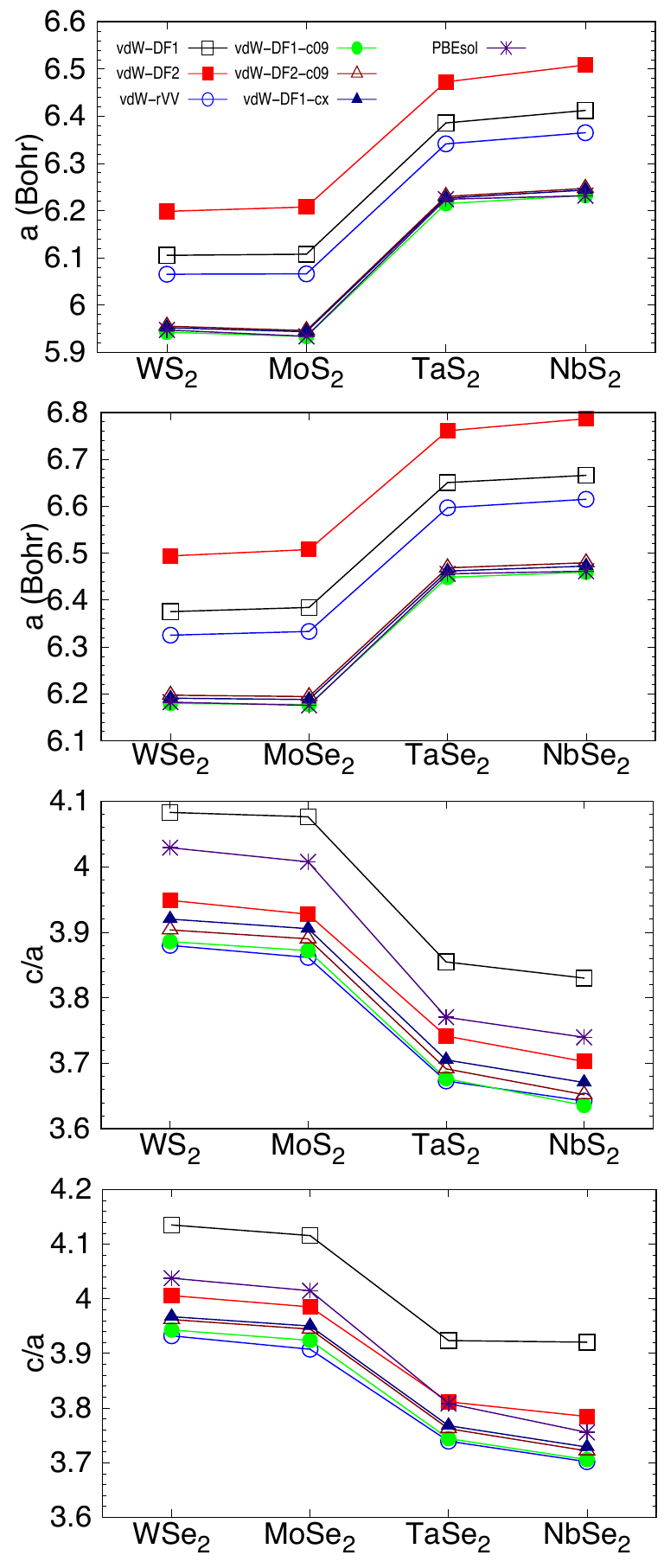}
\caption{\label{lattice} Lattice parameters of MX$_2$, with M = W, Mo, Ta, and Nb, and X = S and Se, calculated using various vdW functionals and PBEsol. The in-plane lattice parameter $a$ is systematically larger in metallic TMDs than in semiconductors, whereas the $c/a$ ratio is smaller in metallic compounds. Connecting lines illustrate the trend between semiconducting and metallic TMDs.}
\end{figure}

The lattice parameters obtained using vdW functionals and the GGA-PBEsol XC functional are presented in Figure~\ref{lattice}. The results agree with the experimental data, confirming that semiconducting TMDs exhibit smaller lattice parameters in the plane ($a$) compared to their metallic counterparts. Moreover, the ratio $c/a$ is consistently smaller for metallic TMDs relative to semiconducting ones, in line with experimental observations. Importantly, these trends are independent of the XC approximation.

The sensitivity of the calculated lattice parameters to the XC functional is evident. Among the tested functionals, vdW-DF2 yields the largest $a$ values, while vdW-DF1 produces the highest $c/a$ in all the TMDs studied (Tables~\ref{a_para}). The lattice parameters obtained from PBEsol, vdW-DF1-c09, vdW-DF2-c09, and vdW-DF1-cx are in close agreement with each other.

Tables~\ref{a_para} highlight the XC functionals that yield the results closest to the experiment. For all MX$_2$ compounds studied, the vdW-DF2-c09 and vdW-DF1-cx functionals provide the most accurate in-plane lattice parameter $a$. Similarly, these functionals also offer the best $c/a$ estimates for WS$_2$ and MoS$_2$. However, for TaS$_2$ and NbS$_2$, the vdW-rVV and vdW-DF1-c09 functionals provide better predictions for $c/a$. For MSe$_2$ compounds, vdW-rVV and vdW-DF1-c09 produce the most accurate $c/a$ values.

The observed differences in lattice parameters between semiconducting and metallic TMDs can be attributed to the fundamental nature of their bonding. In semiconducting TMDs, complete charge transfer from M$^{4+}$ to X$^{2-}$ results in strongly ionic M-X bonds, which contract the in-plane lattice parameter $a$. Meanwhile, interlayer interactions remain weak, allowing for an expanded parameter $c$ and a higher $c/a$ ratio. In contrast, metallic TMDs, where M$^{3+}$ cations exhibit incomplete charge transfer, experience weaker ionic bonding within the layer, leading to an increased $a$ parameter. Furthermore, reduced electron transfer promotes covalent interactions between X atoms both within the layers and across adjacent layers, resulting in a more compact stacking and a smaller $c/a$ ratio. These bonding effects provide a robust explanation for the distinct structural characteristics observed in semiconducting versus metallic TMDs.
 
\begin{table*}
\begin{tabular}{|c|c c c c|c c c c|}
\hline\hline
        &       & $a (Bohr)$&        &         &        &   $c/a$ & & \\
\hline
XC          & WS$_2$ & MoS$_2$ & TaS$_2$ & NbS$_2$ & WS$_2$ & MoS$_2$ & TaS$_2$ & NbS$_2$  \\
\hline
vdW-DF1     & 6.1052 & 6.1077 & 6.3856 & 6.4121& 4.0827 & 4.0762 & 3.8546 & 3.8301  \\
vdW-DF2     & 6.1985 & 6.2076 & 6.4727 & 6.5085& 3.9487 & 3.9273 & 3.7416 & 3.7027 \\
vdW-rVV     & 6.0653 & 6.0661 & 6.3415 & 6.3651& 3.8798 & 3.8615 & {\bf 3.6730} & {\bf 3.6424} \\
vdW-DF1-c09 & 5.9424 & 5.9333 & 6.2148 & 6.2322& 3.8858 & 3.8718 & {\bf 3.6765} & {\bf 3.6359}\\
vdW-DF2-c09 & {\bf 5.9551} & {\bf 5.9458} & {\bf 6.2299} & {\bf 6.2471}& {\bf 3.9035} & {\bf 3.8899} & 3.6916 & 3.6521 \\
vdW-DF1-cx  & {\bf 5.9521} & {\bf 5.9435} & {\bf 6.2265} & {\bf 6.2435}& {\bf 3.9199} & {\bf 3.9055} & 3.7052 & 3.6706\\
PBEsol      & 5.9471 & 5.9338 & 6.2242 & 6.2314& 4.0289 & 4.0074 & 3.7703 & 3.7397\\
Experiment  & 5.9583\cite{Schutte}&5.9715\cite{Bronsema} &6.2644\cite{Givens} &6.2549\cite{Leroux} & 3.9074\onlinecite{Schutte} &3.8892\onlinecite{Bronsema} &3.5982\cite{Givens} & 3.592\cite{Leroux} \\
\hline\hline
XC          & WSe$_2$ & MoSe$_2$ & TaSe$_2$ & NbSe$_2$ & WSe$_2$ & MoSe$_2$ & TaSe$_2$ & NbSe$_2$ \\
\hline
vdW-DF1     & 6.3753 & 6.3844 & 6.6505 & 6.6657& 4.1349 & 4.1156 & 3.9235 & 3.9207\\
vdW-DF2     & 6.4941 & 6.5078 & 6.7609 & 6.7864& 4.0059 & 3.9852 & 3.8115 & 3.7845\\
vdW-rVV     & 6.3253 & 6.3331 & 6.5968 & 6.6147& {\bf 3.9321} & {\bf 3.9075} & {\bf 3.7393} & {\bf 3.7018}\\
vdW-DF1-c09 & 6.1799 & 6.1773 & 6.4486 & 6.4593& {\bf 3.9428} & {\bf 3.9240} & {\bf 3.7441} & {\bf 3.7058} \\
vdW-DF2-c09 & {\bf 6.1974} & {\bf 6.1945} & {\bf 6.4687} & {\bf 6.4795}& 3.9620 & 3.9444 & 3.7624 & 3.7216 \\
vdW-DF1-cx  & {\bf 6.1908} & {\bf 6.1879} & {\bf 6.4618} & {\bf 6.4725}& 3.9671 & 3.9501 & 3.7679 & 3.7286 \\
PBEsol      & 6.1824 & 6.1756 & 6.4561 & 6.4615& 4.0377 & 4.0146 & 3.8082 & 3.7555 \\
Experiment &6.2021\cite{Schutte} &6.2134\cite{Coehoorn} &6.4628\cite{AnYan} &6.5044\cite{Givens}&3.9488\cite{Schutte} &3.9233\cite{Coehoorn} &3.6696\cite{Coehoorn}  &3.6397\cite{Givens} \\
\hline\hline
\end{tabular}
\caption{\label{a_para} The $a$ lattice parameter and $c/a$ ratio of MS$_2$ and MSe$_2$ (M=W, Mo, Ta, Nb) obtained using different XC functionals and compared with experiment. The bold numbers are the closest to the experiment.}
\end{table*}
\begin{figure*}
\includegraphics[width=1.0\textwidth]{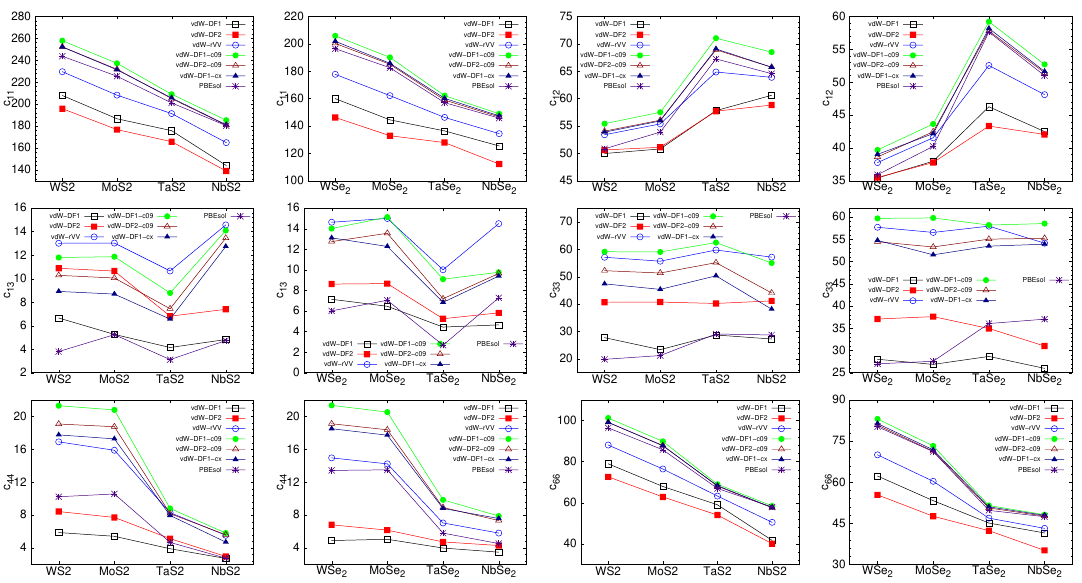}\
\caption{\label{Cij} Elastic constants $c_{ij}$ of MX$_2$ in GPa, with M = W, Mo, Ta, and Nb, and X = S and Se, calculated by different vdW functionals and the PBEsol. } 
\end{figure*}

\subsection{Elastic Constants}
The elastic constants characterize the response of a crystal to applied stress and strain, governing their mechanical stability and anisotropic behavior. In the hexagonal structure of TMDs, there are five independent elastic constants. Figure~\ref{structure}-(b) highlights the most relevant ones, including $c_{11}$ (in-plane normal stiffness), $c_{33}$ (out-of-plane compressive stiffness), $c_{44}$ (out-of-plane shear modulus) and $c_{66}=[c_{11}-c_{12}]/2$ (in-plane shear modulus).

Table~\ref{cijMoS2} compares our DFT-calculated elastic constants for MoS$_2$ with previous theoretical and experimental results. Notably, none of the employed XC functionals predict all $c_{ij}$ values accurately, with vdW-DF1, vdW-DF2, and vdW-rVV exhibiting significant deviations. The data suggest that the local contribution to the Hamiltonian plays a more dominant role in determining $c_{ij}$ than non-local vdW interactions. This trend likely extends to other semiconducting TMDs. Interestingly, the LDA approximation provides better agreement for both diagonal and off-diagonal components of the elastic tensor, reinforcing the minimal impact of vdW interactions on elastic constants.

The trend of changes in $c_{ij}$, obtained using different XC functionals, can be observed from Figure~\ref{Cij}. Analyzing diagonal elastic constants $c_{11}$, $c_{33}$, and $c_{44}$, indicates that metallic MX$_2$ compounds exhibit lower in-plane stiffness $c_{11}$ compared to their semiconducting counterparts. Additionally, replacing S with Se reduces $c_{11}$ for both metallic and semiconducting systems, attributed to the lower electronegativity of Se, which weakens charge transfer and M-X bonding strength. For a given M, the out-of-plane shear constant $c_{44}$ remains nearly unchanged when replacing S with Se, although semiconducting TMDs consistently show higher $c_{44}$ values than metallic ones.

The out-of-plane stiffness constant $c_{33}$ remains nearly invariant across semiconducting and metallic MX$_2$ compounds and is significantly smaller than $c_{11}$. This further confirms the weak interlayer coupling characteristic of these materials, which is consistent with the layered structure of TMDs, where interlayer interactions are predominantly vdW in nature.

\begin{figure*}
\includegraphics[width=1.0\textwidth]{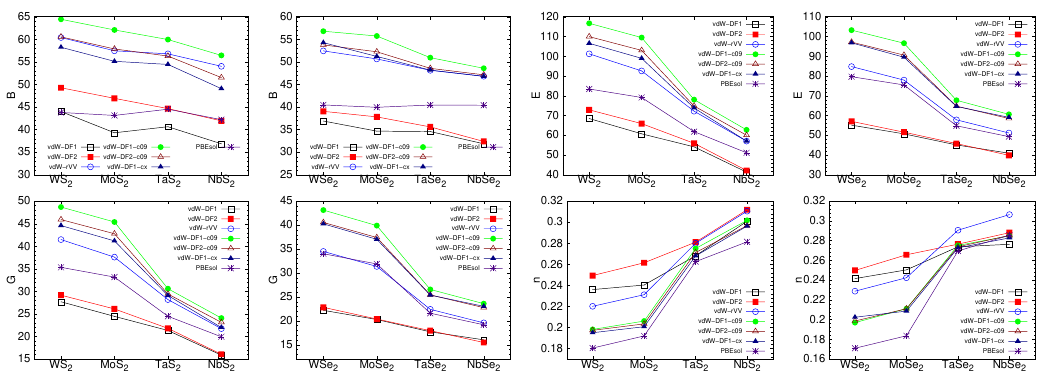}
\caption{\label{modulus} Bulk (B), Young (E), Shear (G)  moduli in GPa and Poisson ratio (n) of MX$_2$, with M = W, Mo, Ta, and Nb, and X = S and Se, calculated by different vdW functionals and the PBEsol. } 
\end{figure*}

The calculated bulk modulus (B), Young’s modulus (E), shear modulus (G), and Poisson’s ratio (n) are presented in Figure~\ref{modulus}. Although their absolute values are strongly dependent on the XC functional, clear trends emerge. Across all XC approximations, metallic MX$_2$ compounds exhibit lower B, E, and G compared to semiconducting TMDs, reflecting their softer mechanical response. Furthermore, increasing the atomic number of transition metal (M) enhances mechanical stiffness, with WX$_2$ and TaX$_2$ displaying higher values of B, E, and G compared to MoX$_2$ and NbX$_2$. These trends align with the periodic table, where heavier transition metals contribute to stronger bonding and greater resistance to deformation.

Comparison of the calculated structural properties of MS$_2$ and MSe$_2$ where M = W, Mo, Ta, and Nb indicates that the bulk moduli B, Young E, and Shear G of the metallic TMDs are smaller than those of the semiconducting structures. Taking into account the electron configuration of the elements M, it can be observed that the moduli B, E, and G of WX$_2$ and TaX$_2$, in which W and Ta belong to period six of the periodic table, are larger than MoX$_2$ and NbX$_2$ whose M ions are located in period five of the periodic table. Hence, increasing the atomic number of M in metallic and semi-conducting MX$_2$ compounds increases the resistance to compression (B), the elastic shear stiffness (E), and the tensile stiffness (G) of TMDs. 

These results highlight the role of chemical composition and bonding in governing the elastic properties of TMDs. The reduced in-plane stiffness and shear modulus in metallic TMDs are particularly relevant for understanding their susceptibility to structural instabilities, such as charge density wave (CDW) formation, further emphasizing the importance of elastic anisotropy in these materials. Our results agree with the previous study in which the first principle method was used to study the elastic behavior of a larger category of TMDs with detailed description of determining the structural modulus \cite{Price2023}.
\begin{table}
\begin{tabular}{|c| c c c c c|}
\hline\hline
method  & $c_{11}$ & $c_{12}$ & $c_{13}$ & $c_{33}$ & $c_{44}$ \\
\hline
vdW-DF1     & 186.6       & 50.8 & 5.2  & 23.3       & 5.4  \\
vdW-DF2     & 176.7       & 51.1 & 10.6 & 40.7       & 7.7  \\
vdW-rVV     & 208.1       & 55.3 & 13.0 & 55.6       & 15.8 \\
vdW-DF1-c09 & {\bf 237.1} & 57.5 & 11.8 & 59.0       & 20.7 \\
vdW-DF2-c09 & 231.7       & 56.1 & 10.0 & {\bf 51.3} & {\bf 18.7} \\
vdW-DF1-cx  & 231.2       & 55.9 & 8.7  & 45.3       & 17.3 \\
PBEsol      & 225.3       & 53.9 & 5.2  & 21.2       & 10.5 \\
HSE06-D2\cite{Peelaers14}    & {\bf 238} & 64       & 12       & 57       & 18\\
GGA\cite{Wei}              & 211       & 49       & 3        & 37       & 30\\
GGA\cite{Todorova}         & 211       & {\bf -62}& {\bf 26} & 42       & {\bf 19}\\
LDA\cite{Todorova}         & 240       & {\bf -63}& {\bf 32} & {\bf 53} & 26\\
periodic HF\cite{Alexiev}  & 255       & -38      & 17       & 35       & 15\\
periodic HF\cite{Todorova} & 218       & -21      & 39       & 59       & 30\\
Experiment\cite{Feldman}   & 238       & -54      & 23       & 52       & 19 \\
\hline\hline
\end{tabular}
\caption{\label{cijMoS2} Elastic constants (GPa) of MoS$_2$ obtained using different methods. The bold numbers are the closest to experiment.}
\end{table}

\subsection{Elastic Anisotropy and Charge Density Wave Formation in Metallic TMDs}
The emergence of CDWs in metallic TMDs is intrinsically linked to their lattice dynamics and elastic properties. In particular, the pronounced reductions in the in-plane stiffness constant $c_{11}$ and the in-plane shear deformation $c_{66}$ observed in metallic TMDs compared to their semiconducting counterparts indicate a more flexible lattice response. In particular, suppression of $c_{11}$ reflects a lower resistance to compressive stress in the plane, while reduction of $c_{66}$ signifies a weakening of the shear rigidity within the basal plane. These factors collectively facilitate the formation of periodic lattice distortions that accompany the CDW state.

A crucial feature of CDW formation in metallic TMDs is the presence of orbitally disparate energy valleys at and around the Fermi level, which allows for strong electron-phonon coupling and drives periodic lattice reconstruction~\cite{lin2020,Rossnagel2011,Zhu2015,Xu2021}. The interplay between this electronic instability and the mechanical softness of the lattice may explain why TaS$_2$ and NbSe$_2$ exhibit well-known 3×3 CDW phases. Specifically, the reduction in $c_{66}$ allows a shear deformation pattern that is consistent with the out-of-phase displacement of metal atoms in the CDW phase~\cite{Xi2015,Ugeda2016}. Simultaneously, the lowered $c_{11}$ facilitates periodic contractions and expansions of the unit cell, further stabilizing the modulated state. These results align with previous studies highlighting that the CDW transition temperature in TMDs is strongly correlated with their elastic anisotropy and soft phonon modes~\cite{Ahn2024}. Furthermore, $c_{33}$, which governs the out-of-plane compressibility, remains largely unchanged across these compounds, highlighting that the CDW formation primarily originates from in-plane lattice instabilities rather than interlayer interactions. This agrees with experiments where similar CDW patterns were obversed in bulk and in the thin film of these materials ~\cite{lin2020,Xi2015}.

The correlation between reduced elastic moduli and CDW formation suggests that metallic TMDs with lower $c_{11}$ and lower $c_{66}$ are more susceptible to spontaneous symmetry-breaking lattice distortions. The results presented in Fig.~\ref{Cij} and Tab.~\ref{cijMoS2} confirm this trend, reinforcing the view that the intrinsic mechanical flexibility of these compounds plays a crucial role in their electronic and structural phase transitions. Given that CDW phenomena significantly impact electronic transport and superconductivity in TMDs, understanding the elastic response provides an essential framework for tuning these materials for future applications in nanoelectronics and quantum devices.

\subsection{Soft phonons and Charge Density Wave Formation in Metallic TMDs}

The onset of CDW order in crystalline materials is closely tied to phonon softening at finite wavevectors. Within linear-response theory, such instabilities are captured when the renormalized phonon frequency $\omega(\mathbf{q})$ becomes imaginary, typically due to the strong electron-phonon coupling at certain $\mathbf{q}$ points. This interaction leads to a renormalization of the phonon dispersion according to $\omega^2(\mathbf{q}) = \omega_0^2(\mathbf{q}) + 2\omega_0(\mathbf{q}) \Re[ \Sigma(\mathbf{q}, \omega)]$, where $\omega_0$ is the bare frequency and $\Sigma$ is the phonon self-energy~\cite{Allen1972,RevModPhys.89.015003}. When the real part of $\Sigma$ is sufficiently negative, $\omega^2(\mathbf{q})$ can turn negative, indicating a dynamical instability of the lattice~\cite{DFPT2001}.

CDW order is an ubiquitous phenomenon among metallic TMDs~\cite{lin2020}. Prototypically, in 2H-NbSe$2$, experimental studies have established a commensurate CDW phase characterized by a wavevector $\mathbf{q}_{\text{CDW}} \approx (1/3, 0, 0)$ in reciprocal lattice units~\cite{Weber2011, Ugeda2016}. This corresponds to a tripling of the unit cell along the in-plane directions. Unlike Peierls-type CDWs in strictly one-dimensional systems, the CDW in NbSe$_2$ emerges in a layered metallic environment with a complex Fermi surface topology, where both nesting and mode-selective electron-phonon coupling play roles~\cite{Weber2011}. Predicting this instability correctly requires an accurate description of both the electronic screening and the long-range dispersive interactions characteristic arising from interlayer interactions.

\begin{figure*}
\includegraphics[width=\textwidth]{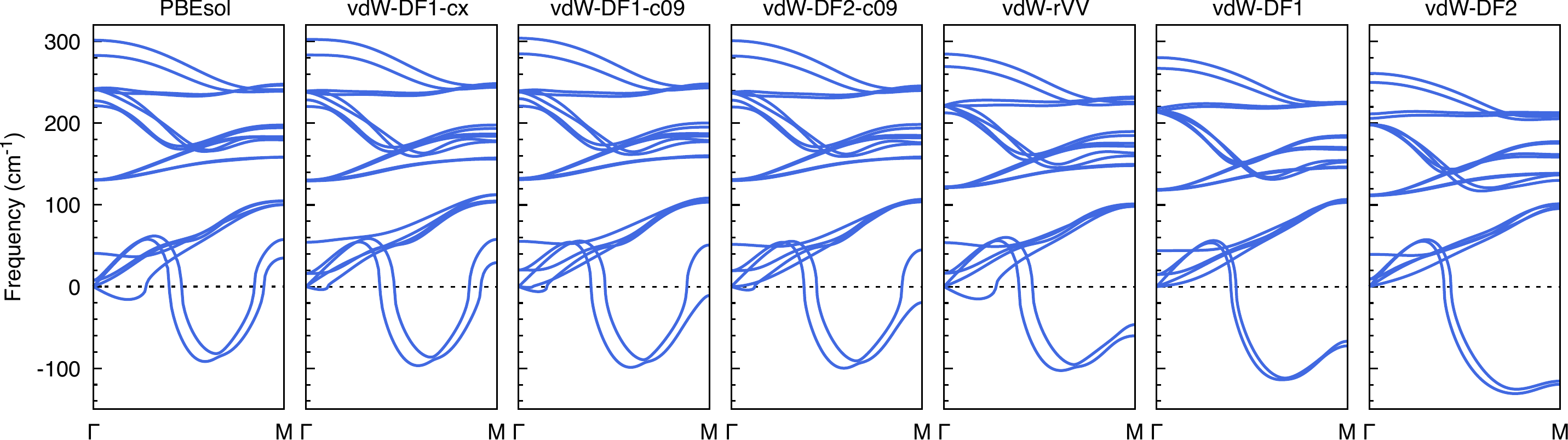}
\caption{\label{Fig5} The calculated phonon band structure of bulk NbSe$_2$ along the $\Gamma - M$ direction with various XC functionals and ordered according to the shift in phonon frequency from positive to negative at the $M$ point.} 
\end{figure*}

To investigate this, we have systematically examined the phonon dispersion of 2H-NbSe$_2$ along the $\Gamma$–M direction using various XC functional, shown in Figure~\ref{Fig5}. As can be seen, all functionals considered predict a dip in the lowest acoustic phonon branch along this path, indicating a lattice instability, consistent with the observed $\mathbf{q}_{\text{CDW}}$. However, the precise location of the soft mode and its associated wavevector vary with the XC functional used.

As shown in Table~\ref{tab:qvec}, functionals such as PBEsol, vdW-DF-cx, and vdW-DF1-c09 predict softening at wavevectors $q <0.3$, thus underestimating the experimentally observed $\mathbf{q}_{\text{CDW}}$. In contrast, the nonlocal correlation functionals vdW-DF1 and vdW-DF2 predict instabilities at larger $q$ values, with vdW-DF1 yielding $q = 0.3275$, in close agreement with the experiment. However, this improved match in q$_{\text{CDW}}$ comes with important caveats. Specifically, both vdW-DF1 and vdW-DF2 exhibit imaginary phonon frequencies not only near $(1/3, 0, 0)$, but extending all the way to the M point, where additional branches also soften. These instabilities at or near the M point imply lattice distortions with different periodicities than the experimentally observed $3 \times 3$ CDW and are not supported by experimental data.

The presence of imaginary frequencies at M, particularly in two branches as seen in vdW-DF1 and vdW-DF2, suggests that these functionals may overestimate the electron-phonon coupling strength or incorrectly model the interatomic force constants at zone boundaries. Such artifacts indicate instabilities beyond the known CDW distortion, and although they signal a strong tendency for the lattice to distort, they do not correspond to the experimentally realized ground state. Thus, while vdW-DF1 and vdW-DF2 better reproduce the location of the CDW vector, they simultaneously predict additional instabilities inconsistent with the known lattice dynamics of NbSe$_2$.

Finally, we note that the correlation between the optimized in-plane lattice constant $a$ and the predicted q$_{\text{CDW}}$ is not straightforward. For instance, vdW-DF-cx and vdW-DF1-c09 yield lattice constants closest to the experiment but do not reproduce the correct CDW wavevector. In contrast, vdW-DF1 produces a significant overestimation in $a$, yet yields a q$_{\text{CDW}}$ value closest to the experimental result. This illustrates that matching structural parameters alone is not sufficient to ensure predictive accuracy for phonon instabilities and that care must be taken in evaluating the performance of XC functionals for lattice-dynamical properties.
\begin{table}[]
    \centering
    \begin{tabular}{c|c|c|c}
    \hline\hline
    Method      & $\omega_{min}~(cm^{-1})$ & q$_{\text{CDW}}$ & $\Delta a~(Bohr)$ \\
    \hline
    PBEsol      & -91.6689   & 0.2928 & -0.0429\\
    vdW-DF-cx   & -96.6969   & 0.2940 & -0.0319 \\
    vdW-DF1-c09 & -98.7498   & 0.2935 & -0.0451 \\
    vdw-DF2-c09 & -99.7767   & 0.2951 & -0.0249\\
    vdW-rVV     & -102.6837  & 0.3063 & 0.1103\\
    vdw-DF1     & -113.9750  & 0.3275 & 0.1613\\
    vdw-DF2     & -130.8664  & 0.3830 & 0.2820\\
    Exp. \onlinecite{Weber2011} & $\cdots$ & 0.328 & $\cdots$ \\
    \hline\hline         
    \end{tabular}
    \caption{Minimum of acoustic branch in phonon spectra (Fig. \ref{Fig5}) $\omega_{min}$, and value of charge density vector $q_{CDW}$ obtained by different XC approximation. The last column shows the difference between lattice parameter $a$ which is obtained by different XC approximation and experiment: $\Delta a = a_{\text{XC}} - a_{\text{exp}}$. }
    \label{tab:qvec}
\end{table}

Anharmonic effects were not considered in our phonon calculations. These effects can suppress CDWs in metallic TMDs \cite{Leroux}, and this is an important part of understanding why not all soft phonon modes lead to actual CDW transitions, even when they appear unstable in harmonic phonon calculations. The anharmonicity can stabilize modes that appear imaginary in the harmonic approximation because the potential energy surface may have a shallow double well or be flat, meaning atoms can vibrate around without a real phase transition, and therefore considering cubic or even quadratic terms in potential energy surface expansion becomes crucial.

\section{Conclusion} 
In this work, we performed a comprehensive DFT and DFPT study of the structural, elastic properties, and phonon spectra of various 2H-MX$_2$ compounds using various vdW exchange-correlation functionals. Our analysis revealed distinct trends in lattice parameters and elastic moduli, which differentiate semiconducting TMDs from metallic TMDs. Metallic compounds were shown to exhibit larger in-plane lattice parameters ($a$) and reduced interlayer spacing ($c/a$), consistent with their weaker ionic bonding and enhanced covalent interactions between adjacent layers.

By analyzing the elastic constants, we found that metallic TMDs exhibit significantly lower in-plane stiffness ($c_{11}$) and shear modulus ($c_{66}$) compared to their semiconducting counterparts. We discussed the correlation between this mechanical softening and the emergence of charge density wave states observed in these systems. By benchmarking DFPT results for phonon spectra of NbSe$_2$, we compared the predicted values for q$_\text{CDW}$ obtained by studied XC approximations. 

Our study also examined the accuracy of different exchange-correlation functionals, identifying vdW-DF2-c09 and vdW-DF1-cx as the most reliable for lattice parameters, while vdW-rVV and vdW-DF1-c09 best predict elastic constants. We found that vdW-DF1 provides the best value for q$_\text{CDW}$ of NbSe$_2$ compared to experiment. These insights are crucial for modeling mechanical and physical properties in layered materials.

The strong dependence of elastic properties on chemical composition and electronic structure suggests that TMDs can be mechanically tailored for applications in flexible electronics and quantum materials. Additionally, the connection between elastic softness and CDW formation provides new opportunities to tune phase transitions through strain engineering or substrate interactions.

\section{Acknowledgment}
We acknowledge the support of the Leverhulme Trust under Grant Agreement No. RPG-2023-253. R.V.B. and M.S.B. are grateful to E-IMR center at the Institute for Materials Research, Tohoku University, for continuous support. S. Azadi and T.D. K\"{u}hne acknowledge the computing time provided to them on the high-performance computers Noctua2 at the NHR Center in Paderborn (PC2). A.A. received support from the Indonesia Endowment Fund for Education (LPDP), NIB/202209223311735. A.A. gratefully acknowledges the Research Facility at MAHAMERU BRIN HPC under the National Research and Innovation Agency of Indonesia.
\bibliography{mainbib}

\end{document}